\documentclass[prd,showpacs,twocolumn,preprintnumbers,amsmath,amssymb,superscriptaddress,floatfix,nofootinbib]{revtex4-2}%

\usepackage{graphicx}
\usepackage{bm}
\usepackage{amsmath}
\usepackage{amsfonts}
\usepackage{amssymb}
\usepackage{color}
\usepackage{multirow}
\usepackage{subfigure}
\usepackage{epstopdf}
\usepackage{appendix}
\usepackage[colorlinks, citecolor=blue,anchorcolor=red,menucolor=red, linkcolor=red,filecolor=red,runcolor=red,urlcolor=blue,frenchlinks=red]{hyperref}

\begin{document}
	\title{Identifying $\Sigma(1380)$ and $\Sigma(1430)$ in the $J/\psi \to \Lambda \pi \bar{\Sigma}$ reaction}

	\author{Wen-Tao Lyu}\email{lvwentao9712@163.com}
	\affiliation{School of Physics, Zhengzhou University, Zhengzhou 450001, China}
	\affiliation{Departamento de Física Teórica and IFIC, Centro Mixto Universidad de Valencia-CSIC Institutos de Investigación de Paterna, 46071 Valencia, Spain}
	\vspace{0.5cm}
	
	\author{Eulogio Oset}\email{oset@ific.uv.es}
	\affiliation{Departamento de Física Teórica and IFIC, Centro Mixto Universidad de Valencia-CSIC Institutos de Investigación de Paterna, 46071 Valencia, Spain}
	\vspace{0.5cm}
	
	\author{De-Min Li}\email{lidm@zzu.edu.cn}
	\affiliation{School of Physics, Zhengzhou University, Zhengzhou 450001, China}\vspace{0.5cm}
 
	\author{En Wang}\email{wangen@zzu.edu.cn}
	\affiliation{School of Physics, Zhengzhou University, Zhengzhou 450001, China}\vspace{0.5cm}

\begin{abstract}
We study the $J/\psi \to \Lambda \pi \bar{\Sigma}$ reaction by looking at the $\pi^+ \Lambda$ mass distribution at low energies, in search of signals for the low lying $\Sigma^+$ excited states. Apart from a clear signal of the $\Sigma(1385) (3/2^+)$ state, we find a smaller peak for the predicted $\Sigma(1430) (1/2^-)$, which has already been confirmed by the Belle Collaboration. A first analysis, considering only the $\pi\Lambda$ interaction, shows that the low energy part of the spectrum is better reproduced including contributions from the $\Sigma(1430)$ and the predicted $\Sigma(1380)(1/2^-)$ state that has been claimed before from analyses of different experiments. However, when we consider the $\pi\bar{\Sigma}$ interaction the need for the $\Sigma(1380)$ disappears.

\end{abstract}
	
	%\keywords{ }
	\date{\today}
	
	\maketitle
	
\section{Introduction}\label{sec1}

The $\Sigma^*$ states are the subject of continuous discussions concerning their existence and properties~\cite{Wang:2024jyk}. In this work we shall address two of them which are still object of much discussion at present, the $\Sigma(1430) (J^P=1/2^-)$ and the $\Sigma(1380) (J^P=1/2^-)$. The $\Sigma(1430)$ was predicted theoretically in the study of the meson baryon interaction in the context of the chiral unitary approach~\cite{Oller:2006jw}, and it appeared as a cusp in the $\bar{K}N$ threshold, or as a slightly bound state depending of some small differences in the input of the interaction. Those results are corroborated by further studies imposing constraints from $\pi N$ and $\bar{K}N$ data~\cite{Lu:2022hwm}. Other works along the lines of the chiral unitary approach in coupled channels come to support these findings~\cite{Oset:2001cn,Khemchandani:2018amu,Kamiya:2016jqc,Oller:2006jw,Garcia-Recio:2002yxy,Lutz:2001yb,Guo:2012vv}, and the $I = 1$ state around the $\bar{K}N$ threshold remains when the mixing of pseudoscalar-baryon and vector-baryon components is implemented~\cite{Khemchandani:2012ur,Khemchandani:2011mf}. The data from $\pi \Sigma$ photoproduction in $\gamma p \to K\Sigma\pi$~\cite{CLAS:2013rjt} were considered in Ref.~\cite{Roca:2013cca} to constrain the parameters of the theory and, again, a cusp-like structure in $|T|^2$ appeared for the different amplitudes at the $\bar{K}N$ threshold.

To search for the $\Sigma(1430)$, many reactions have been proposed. In Ref.~\cite{Lyu:2023oqn} the $\gamma n \to K^+\Sigma(1430)^{-}$ reaction was proposed to observe this state. In Refs.~\cite{Ren:2015bsa,Wu:2013kla} the $\bar{\nu}_l p \to l^+ \phi B$ reaction, with $\phi$ and $B$ being mesons and baryons of the SU(3) octet, was also shown to be appropriate to observe this state. It was proposed in Ref.~\cite{Wang:2015qta} to search for the $\Sigma(1430)$ in the $\Sigma\pi$ mass distribution of the process $\chi_{c0}(1P) \to \Sigma\bar{\Sigma}\pi$. A similar work suggested using the $\chi_{c0} \to \Lambda\bar{\Sigma}\pi$ decay, considering the contributions from the $\pi\Sigma$ and $\pi\Lambda$ final state interactions~\cite{Liu:2017hdx}. In Ref.~\cite{Xie:2018gbi}, the $\Lambda_c^+ \to \pi^+\pi^0\pi^-\Sigma^+$ reaction was analyzed in terms of a triangle singularity that reinforced the production of the $\Sigma(1430)$ state. Also, in Ref.~\cite{Xie:2018gbi}, it was suggested to look at the $\Lambda_c^+ \to \pi^+\pi^+\pi^-\Lambda$ reaction, where the signal for this resonance should be equally seen. The Belle Collaboration measured $\Lambda_c^+ \to \pi^+\pi^+\pi^-\Lambda$ recently~\cite{Belle:2022ywa} and found a clear signal at the $\bar{K}N$ threshold, both in the $\Lambda\pi^+$ and $\Lambda\pi^-$ mass distributions, providing the first clear evidence for the existence of this state. It is also shown that one cannot discriminate between having a resonance or a cusp at the $\bar{K}N$ threshold~\cite{Belle:2022ywa}. Yet, from our point of view, there is only a small and smooth transition from one scenario to the other by gradually changing the strength of the interaction. The $\Lambda_c^+ \to \pi^+\pi^+\pi^-\Lambda$ reaction was studied later in Ref.~\cite{Li:2025yad} with the same framework of Ref.~\cite{Xie:2018gbi}, and it was found that the signal and production rates of the $\Sigma(1430)$ are in good agreement with the Belle results. 

The other state, $\Sigma(1380)$, is still more elusive. Its mass is around that of the $\Sigma(1385)(J^P=3/2^+)$ which shows up overwhelmingly in most reactions in the $\pi\Lambda$ mass distribution, making the identification of $\Sigma(1380)$ difficult. But the $\Sigma(1380)$ was suggested in Ref.~\cite{Wu:2009nw} as a means to improve the agreement with the data of the $K^- p \to \Lambda \pi^- \pi^+$ reaction of Ref.~\cite{Mast:1973gb}. One thing that one should make clear from the beginning is that the $\Sigma(1380)$ cannot have a molecular structure as the $\Sigma(1430)$. The reason is that all the $\Sigma(1/2^-)$ states stemming from the interaction of meson baryon coupled channels were already investigated in Ref.~\cite{Jido:2003cb}, and there were only two isospin $I=1$ states, one the mentioned $\Sigma(1430)$ and another one at higher energies that was identified with the $\Sigma(1620)(1/2^-)$. There was no room for more $I=1$ states with $1/2^-$. However, there could be states with other structures that we cannot rule out. In that sense in Ref.~\cite{Zhang:2004xt} it is shown that ordinary three quarks states in the quark model would have masses even higher than $1430 \text{ MeV}$, but some particular pentaquark structures could have a mass compatible with $1380 \text{ MeV}$. Similar conclusions are reached in Ref.~\cite{Yao:2025qor}. 

Further support for the $\Sigma(1380)$ state has come from different works. In Ref.~\cite{Gao:2010hy} the contribution of $\Sigma(1380)$ was shown to improve the agreement with the LEPS data of the $K\Sigma^*$ photoproduction~\cite{LEPS:2008azb}. Following that line of research, it was shown in Ref.~\cite{Chen:2013vxa} that, using linearly polarized beam in the $\gamma N \to K^+ \Sigma(1385) \to K^+ \pi \Lambda$ reaction near threshold, the signal of the $\Sigma(1380)$ should be further stressed. Similarly, in Ref.~\cite{Xie:2014zga} the agreement with the data on the $\Lambda p \to \Lambda p \pi^0$ reaction was shown to improve with the addition of the $\Sigma(1380)$ contribution. In Ref.~\cite{Xie:2017xwx} the $\Lambda_c^+ \to \eta \pi^+ \Lambda$ decay was studied and it was proposed to study the angle and energy distributions of the pions to differentiate between the $\Sigma(1385)$ and $\Sigma(1380)$ resonances. In Ref.~\cite{Wang:2024ewe} it was suggested to look into the $\Lambda_c^+ \to \gamma \pi^+ \Lambda$ reaction, since a triangle singularity enhanced the role of the $\Sigma(1380)$ contribution. Further support for the $\Sigma(1380)$ was found from the study of the $\Lambda_c \to \eta \pi \Lambda$ reaction in Ref.~\cite{Lyu:2024qgc}. This reaction has been further addressed in Ref.~\cite{Duan:2024czu}, improving in some points, and showing a good agreement with the data of Ref.~\cite{BESIII:2024mbf}, with some clear disagreement at low energies in the $\pi^+ \Lambda$ mass distribution. This failure has stimulated further work in Ref.~\cite{Lyu:2026ack}, attributing the missing strength at low $\pi\Lambda$ invariant masses to the missing contribution of the $\Sigma(1380)$ state. Further work supporting the role of this state is done in Ref.~\cite{He:2025vij} in the study of the $\Lambda_c^+ \to \Lambda \pi^+ \pi^+ \pi^-$ reaction, showing that the inclusion of the $\Sigma(1380)$ resonance significantly improves the fit in the $\pi\Lambda$ low-energy region. Similarly, the spectra of the Belle experiment on the $\Lambda_c^+ \to p K_S^0 \eta$~\cite{Belle:2022pwd} was analyzed in Ref.~\cite{Li:2024rqb}, showing that a fit to the data was improved by including the contribution of a $\Sigma^*(1/2^-)$ state at $1380 \text{ MeV}$. 

In the present work, we study the $J/\psi \to \Lambda \pi \bar{\Sigma}$ reaction, in which the $\Sigma(1385)$ resonance plays a dominant role. We show, however, that signals associated with the two $1/2^-$ states could also be visible in the $\pi\Lambda$ spectrum. We then analyze the reaction by looking first at the $\pi\Lambda$ interaction alone and then also including the $\pi\bar{\Sigma}$ interaction. We find that, while considering only the $\pi\Lambda$ interaction, a contribution from the $\Sigma(1380)$ state is welcome, the need for this resonance disappears when the $\pi\bar{\Sigma}$ interaction is considered in addition.

\section{Formalism}\label{sec2}

\begin{figure}[htbp]
	\centering
	
	%	\centering
	\includegraphics[scale=0.45]{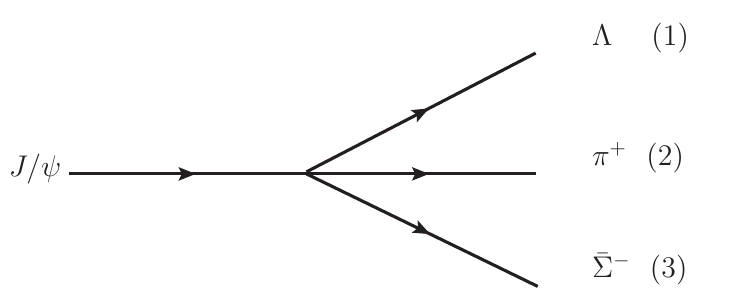}
	
	\caption{Diagrammatic representation of the $J/\psi\to\Lambda\pi^+\bar{\Sigma}^-$ decay.}\label{fig:jpsitoLampiSig}
\end{figure}

We present here the formalism for the $J/\psi \to \bar{\Lambda} \pi \Sigma$, or equivalently $J/\psi \to \Lambda \pi \bar{\Sigma}$ reaction. The decays of charmonium vector mesons into a meson and a pair of baryon antibaryon have been studied theoretically in Refs.~\cite{He:2026mkf,Dai:2026zqn,Lyu:2026rsm}. The common thing between the approaches in these works is the consideration of the $J/\psi$~($c\bar{c}$) as an SU(3) singlet in the $u, d, s$ quarks, which allows to relate the weights for production of different baryon-antibaryon-meson trios. In Refs.~\cite{He:2026mkf,Lyu:2026rsm} an explicit use of the antibaryon spinors is made, while in Ref.~\cite{Dai:2026zqn}, a simplified formalism is used, adopting a $P$-wave coupling in the production of a nonrelativistic antiparticle. Given that the differences in the results with both methods are very small, we adopt the second method in the present work, which makes the formalism more amenable.

The coupling assumed in Ref.~\cite{Dai:2026zqn} for the process of Fig.~\ref{fig:jpsitoLampiSig} is given by
\begin{equation}\label{eq:t}
t = -\tilde{t} \, \epsilon_\mu p^\mu_{\bar{\Sigma}^-},   
\end{equation}
where $\tilde{t}$ carries the information related to the $\pi^+ \Lambda$ interaction and the resonances coupled to this channel. The amplitude $\tilde{t}$ in Eq.~\eqref{eq:t} is a function of the invariant mass $M_{12}$ of the particles 1 and 2, with the labels reflecting the association to each particle in Fig.~\ref{fig:jpsitoLampiSig}.

From the Review of Particle Physics (RPP)~\cite{ParticleDataGroup:2024cfk} formula in the Mandl \& Shaw normalization of the fields~\cite{Mandl:1985bg}, we have:
\begin{equation}\label{eq:double_width}
\frac{d^2\Gamma}{dM_{12}^2 dM_{23}^2} = \frac{1}{(2\pi)^3} \frac{2M_\Lambda 2M_{\bar{\Sigma}}}{32 m_{J/\psi}^3} \overline{\sum}\sum|t|^2,
\end{equation}
where $|t|^2$ is summed over final and averaged over initial polarizations. We need only two independent invariant masses according to the relationship
\begin{equation}\label{eq:M_relation}
M_{12}^2 + M_{13}^2 + M_{23}^2 = m_{J/\psi}^2 + m_{\bar{\Sigma}^-}^2 + m_{\pi^+}^2 + m_\Lambda^2. 
\end{equation}

In our first analysis of the data by looking at the $\pi^+\Lambda$ interaction, it is useful to use other variables, to differentiate between $S$-wave and $P$-wave resonances in the $\pi^+\Lambda$ variables. For this reason, we show in Appendix~\ref{sec:appendix} the equivalence of the formula of Eq.~\eqref{eq:double_width} with another one, where the variables are $M_{12}$ and $\cos\theta$, with $\theta$ the angle between particle 2 ($\pi^+$) with particle 3 ($\bar{\Sigma}^-$) in the $\pi^+\Lambda$ rest frame\footnote{Note that the direction of $\bar{\Sigma}^-$ momentum in the $J/\psi$ rest frame and in the $\pi^+\Lambda$ rest frame are the same, since the boost is done in the $\vec{p}_{\Sigma^-}$ direction.}. Given the independence on the $\phi$ angle in that frame, we take the solid angle $\tilde{\Omega}$ in that frame, and then $d\tilde{\Omega} = 2\pi d\cos\theta$. Using the result in Appendix~\ref{sec:appendix}, we find
\begin{equation}\label{eq:gamma_mass_angle}
\frac{d^2\Gamma}{dM_{12} d\tilde{\Omega}} = \frac{1}{(2\pi)^4} \frac{1}{8 m_{J/\psi}^2} p_{\bar{\Sigma}} \tilde{p}_\pi 2M_{\bar{\Sigma}} 2M_\Lambda \overline{\sum}\sum |\tilde{t}|^2, 
\end{equation}
where $p_{\bar{\Sigma}^-}$ is the $\bar{\Sigma}^-$ momentum in the $J/\psi$ rest frame and $\tilde{p}_{\pi}$ is the $\pi^+$ momentum in the $\pi^+\Lambda$ rest frame, which is the formula used in Ref.~\cite{Bayar:2022wbx}.

Next we look into the mass distribution of the BESIII experiment~\cite{BESIII:2023syz} (see Fig. 3 $(d,I)$ of that reference), and we observe a very strong peak in the $\pi^+ \Lambda$ mass distribution that corresponds to the excitation of the $\Sigma(1385) (3/2^+)$ resonance. A hint of a very small contribution around 1430~MeV can also be observed, and the tail of the $\Sigma(1385)$ resonance at low invariant masses deviates from a symmetrical Breit-Wigner form, which could hint to some extra contribution in that region.

We are concerned about the $\cos\theta$ dependence of the expected amplitudes. The $\Sigma(1385)$ resonance proceeds via $P$-wave in $\cos\theta$, but we also have a tree level contribution in $S$-wave in the angle $\theta$ and the possible signal for the $\Sigma(1430)$ also in $S$-wave in that angle. Hence, we can write
\begin{equation}\label{eq:a_b_angle}
\tilde{t} = C \left[a Y_{00}\left(\cos\theta\right) + b Y_{10}\left(\cos\theta\right)\right]\vec{\epsilon} \cdot \vec{p}_{\bar{\Sigma}^-},
\end{equation}
with $Y_{lm}$ spherical harmonics and $\vec{\epsilon}$ the $J/\psi$ polarization vector. We parametrize
\begin{equation}
\begin{aligned}\label{eq:a_and_b}
a = 1 + \frac{\tilde{a} M_{\Sigma(1430)}}{M_{12} - M_{\Sigma(1430)} + i \Gamma_{\Sigma(1430)}/2}, \\
b = \frac{\tilde{b} \tilde{p}_\pi M_{\Sigma(1385)} / p_{\bar{\Sigma}}}{M_{12} - M_{\Sigma(1385)} + i \Gamma_{\Sigma(1385)}/2} .
\end{aligned}
\end{equation}

Note that in Eq.~\eqref{eq:a_b_angle} we show the explicit angular dependence for the $S$- and $P$-waves, but these terms are accompanied by momentum dependent factors. Note also that we have in mind to use Eq.~\eqref{eq:gamma_mass_angle}, where the variables are $M_{12}$ and the angle $\theta$. Hence the accompanying momentum dependent factors should be functions of $M_{12}$ as we show below.
In the term $b$ of Eq.~\eqref{eq:a_and_b} we have put in the numerator $\tilde{p}_\pi / p_{\bar{\Sigma}}$ (the factor $M_{\Sigma(1385)}$ is put there for dimensional reasons). The reason for this factor is that, unlike the production of $\Sigma(1430)$ via the vertex of Eq.~\eqref{eq:t}, we produce now $\Sigma(1385)$. The amplitude goes as $\vec{S} \cdot \vec{\tilde{p}}_\pi \vec{S}^\dagger \cdot \vec{\epsilon}$, where $\vec{S}^\dagger$ is the transition spin operator from spin 1/2 to 3/2. Thus, instead of a factor $\vec{p}_{\bar{\Sigma}}$ in Eq.~\eqref{eq:t} we have now the factor $\tilde{p}_\pi$. Note that $p_{\bar{\Sigma}}$ and $\tilde{p}_\pi$ are both functions of $M_{12}$, as 
\begin{equation}
\begin{aligned}\label{eq:psigma_and_ppi}
p_{\bar{\Sigma}} = \frac{\lambda^{1/2}(m_{J/\psi}^2, M_{\bar{\Sigma}}^2, M_{12}^2)}{2 m_{J/\psi}}, \\
\tilde{p}_\pi = \frac{\lambda^{1/2}(M_{12}^2, m_{\pi^+}^2, M_{\Lambda}^2)}{2 M_{12}}.
\end{aligned}
\end{equation}
The term 1 in Eq.~\eqref{eq:a_and_b} stands for the tree level contributions with no resonance production and we play a bit with the peak position of the $\Sigma(1430)$ by taking $M_{\Sigma(1430)}=1430\pm10$~MeV and $\Gamma_{\Sigma(1430)}=22\pm3$~MeV, while we take $M_{\Sigma(1385)}=1382.83~\text{MeV}$ and $\Gamma_{\Sigma(1385)}=36.2~\text{MeV}$ from Ref.~\cite{ParticleDataGroup:2024cfk}. The explicit consideration of the $\tilde{p}_{\pi} / p_{\bar{\Sigma}^-}$ factor in Eq.~\eqref{eq:a_and_b} and the average over the $J/\psi$ polarizations in $\frac{1}{3} \sum_{pol} \epsilon_i \epsilon_j = \frac{1}{3} \delta_{ij}$ in the sum and average over polarizations in $|t|^2$, allows us to write
\begin{equation}
\frac{d^2\Gamma}{dM_{12} d\tilde{\Omega}} = \frac{1}{(2\pi)^4} \frac{1}{8 m_{J/\psi}^2} p_{\bar{\Sigma}} \tilde{p}_\pi 2M_{\bar{\Sigma}} 2M_\Lambda \frac{1}{3} \vec{p}_{\bar{\Sigma}^-}^{~2} |\tilde{t}|^2 .
\end{equation}
There is a spin dependence in the $\vec{S} \cdot \tilde{p}_{\pi} \vec{S} \cdot \vec{p}_{\bar{\Sigma}^-} = \frac{2}{3} \tilde{p}_{\pi} \cdot \vec{p}_{\bar{\Sigma}^-} - i \frac{1}{3} \epsilon_{ijk} \sigma_k \tilde{p}_{\pi i} p_{\bar{\Sigma}^- j}$ operator, but since the tree level does not have spin dependence it does not interfere with the spin dependent term and also not with the spin independent term of the $\Sigma(1385)$, as a consequence of which, one can simplify the formalism accommodating the spin dependent contribution in the effective $b$ coefficients. Then, we have three free parameters to fit to the data of $d\Gamma/dM(\pi\Lambda)$: $C$, $\tilde{a}$, $\tilde{b}$.

\section{Results with only the $\pi^+\Lambda$ interaction}\label{sec3}

\begin{figure}[htbp]
	\centering
	
	%	\centering
	\includegraphics[scale=0.65]{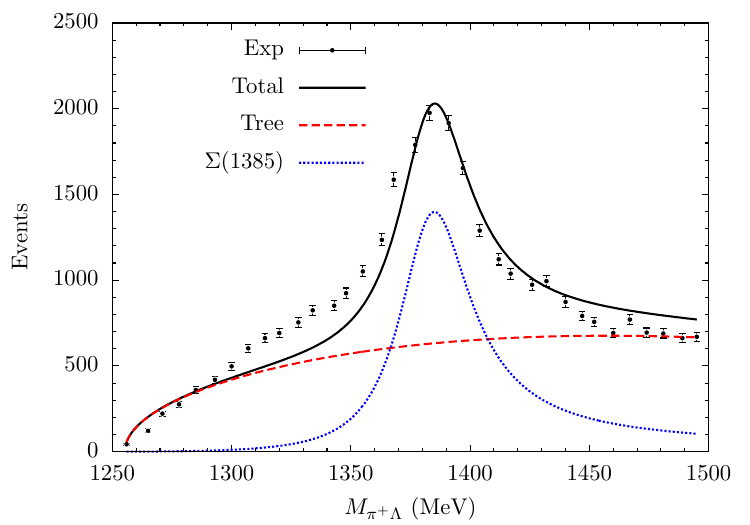}
	
	\caption{$\pi^+\Lambda$ mass distribution without the two $\Sigma(1/2^-)$. The data are taken from Ref.~\cite{BESIII:2023syz}.}\label{fig:2}
\end{figure}
\begin{figure}[htbp]
	\centering
	
	%	\centering
	\includegraphics[scale=0.65]{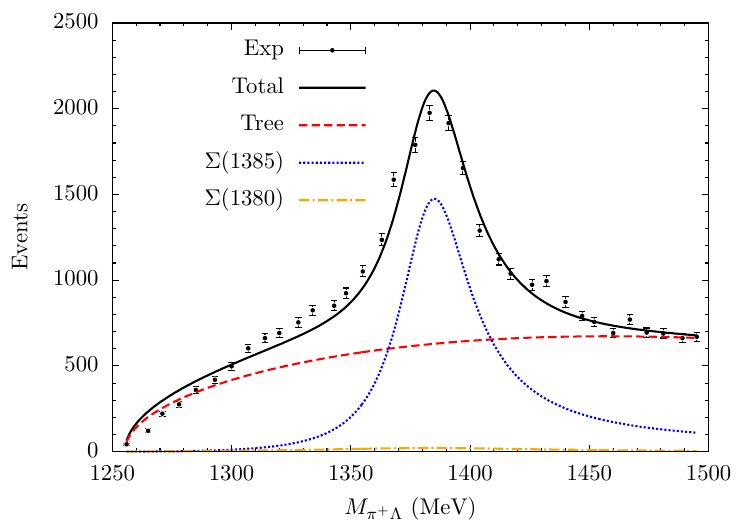}
	
	\caption{$\pi^+\Lambda$ mass distribution with the contribution from $\Sigma(1380)$. The data are taken from Ref.~\cite{BESIII:2023syz}.}\label{fig:3}
\end{figure}
\begin{figure}[htbp]
	\centering
	
	%	\centering
	\includegraphics[scale=0.65]{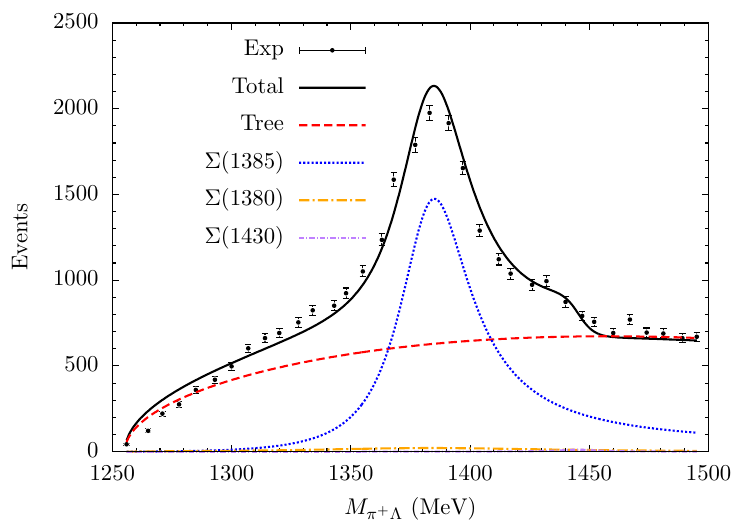}
	
	\caption{The $\pi^+\Lambda$ mass distribution with contributions from $\Sigma(1380)$ and $\Sigma(1430)$. The data are taken from Ref.~\cite{BESIII:2023syz}.}\label{fig:4}
\end{figure}

In Fig.~\ref{fig:2} we plot now the results concerning the input of Eq.~\eqref{eq:a_and_b} taking $C$ and $\tilde{b}$ with $\tilde{a}=0$. We observe that the $\Sigma(1385)$ peak is reasonably reproduced, but there is missing strength at low $\pi^+\Lambda$ invariant mass.

The missing strength in the low energy region of $M_{\text{inv}}(\pi^+\Lambda)$ could be a signal of the missing contribution of the $\Sigma(1380)(1/2^-)$. To substantiate this hint, we replace the $\Sigma(1430)$ resonance in Eq.~\eqref{eq:a_and_b} by the resonance $\Sigma(1380)$ and write
\begin{equation}
a = 1 + \frac{\tilde{a}_1 M_{\Sigma(1380)}}{M_{12} - M_{\Sigma(1380)} + i \Gamma_{\Sigma(1380)}/2}, 
\end{equation}
where, from Refs.~\cite{Wu:2009nw,Wu:2009tu,Liu:2017hdx,Li:2024rqb,Lyu:2024qgc}, we take $M_{\Sigma(1380)} = 1380$~MeV and $\Gamma_{\Sigma(1380)} = 120$~MeV.

What we see in Fig.~\ref{fig:3} is that now we obtain a very good fit to the data. The parameters obtained are $C=3.28\times10^{-4}$, $\tilde{a}_1=-7.99\times10^{-3}$ and $\tilde{b}=-8.29\times10^{-2}$. In the region of $1430 \text{ MeV}$ there is a small peak in the data where we have also a small missing strength, which could be associated to the $\Sigma(1430)$ resonance.

Finally, we add in Eq.~\eqref{eq:a_and_b} the contributions from the $\Sigma(1430)$ and $\Sigma(1380)$. The results are shown in Fig.~\ref{fig:4} by means of the fitted value $\tilde{a}=-9.0\times10^{-4}$. We see now that we obtain a good fit to all the data, filling the missing strength of Fig.~\ref{fig:3} around $1430 \text{ MeV}$.

\section{Consideration of the $ \pi \bar{\Sigma}^- $ interaction}\label{sec4}

While the exercise done above makes the idea of the need of a $ \Sigma(1380) $ state appealing, before making strong claims on it we would like to make further considerations: We would like to see if the $ \pi \bar{\Sigma} $ interaction, which we have not taken into account, could be responsible for the missing strength in the low energy side of the $ \pi^+ \Lambda $ spectrum. This requires a more elaborate formalism that we take from Ref.~\cite{Dai:2026zqn}. In this work the $ J/\psi \to \Lambda \bar{\Sigma}^0\eta$ reaction, which violates isospin, is studied and a good description of the BESIII measurements~\cite{BESIII:2023syz} is obtained.

The starting point in Ref.~\cite{Dai:2026zqn} is to take the $ J/\psi$ state as an SU(3) singlet in the $ u$, $d$, $s $ quarks and then see the combinations of a baryon, an antibaryon and one meson that form an SU(3) singlet. This is accomplished with two structures $ \langle \bar{B} B P \rangle $ and $ \langle \bar{B} P B \rangle $, to which we give weights $ \frac{\tilde{A}}{2} $ and $ \frac{\tilde{B}}{2} $, respectively, where $ B$, $\bar{B} $ and $ P $ are respectively the SU(3) matrices for these hadrons given by

\begin{eqnarray}\label{eq:Pmatrix}
   P=
    \left(
    \begin{array}{ccc}
    \frac{1}{\sqrt{2}}\pi^0 + \frac{1}{\sqrt{3}} \eta & \pi^+ & K^+ \\[2mm]
    \pi^- & -\frac{1}{\sqrt{2}} \pi^0 + \frac{1}{\sqrt{3}} \eta & K^0 \\[2mm]
    K^- & \bar{K}^0 & ~-\frac{1}{\sqrt{3}} \eta\\
    \end{array}
    \right)\,, \quad
\end{eqnarray}
\begin{eqnarray}
%\label{eq:Bmatrix}
   B=
    \left(
    \begin{array}{ccc}
    \frac{1}{\sqrt{2}}\Sigma^0 + \frac{1}{\sqrt{6}} \Lambda & \Sigma^+ & p\\[2mm]
    \Sigma^- & -\frac{1}{\sqrt{2}} \Sigma^0 + \frac{1}{\sqrt{6}} \Lambda & n \\[2mm]
    \Xi^- & \Xi^0 & ~-\frac{2}{\sqrt{6}} \Lambda\\
    \end{array}
    \right),
\end{eqnarray}
\begin{eqnarray}
%\label{eq:Bbarmatrix}
   \bar{B}=
    \left(
    \begin{array}{ccc}
    \frac{1}{\sqrt{2}}\bar{\Sigma}^0 + \frac{1}{\sqrt{6}} \bar{\Lambda} & \bar{\Sigma}^+ & \bar{\Xi}^+ \\[2mm]
    \bar{\Sigma}^- & -\frac{1}{\sqrt{2}} \bar{\Sigma}^0 + \frac{1}{\sqrt{6}} \bar{\Lambda }&  \bar{\Xi}^0\\[2mm]
    \bar{p} & \bar{n} & ~-\frac{2}{\sqrt{6}} \bar{\Lambda}\\
    \end{array}
    \right).
\end{eqnarray}
We do not need all the terms coming from there, since we are interested in the $ \pi \bar{\Sigma} $, and other coupled channels that can lead to $ \pi\bar{\Sigma} $ through final state interaction, we just take the terms that contain a $ \Lambda $ field, and have
\begin{equation}\label{eq:BBP}
\begin{aligned}
 &\langle \bar{B} B P \rangle : \\
 &\frac{\Lambda}{\sqrt{6}} \left\{ \pi^0 \bar{\Sigma}^0 + \pi^+ \bar{\Sigma}^- + \pi^- \bar{\Sigma}^+ - \frac{\sqrt{2}}{3} \eta \bar{\Lambda} + K^+ \bar{p} + K^0 \bar{n} \right.\\
&\left.- 2K^- \bar{\Xi}^+ - 2\bar{K}^0 \bar{\Xi}^0 \right\}, 
\end{aligned}
\end{equation}
\begin{equation}\label{eq:BPB}
\begin{aligned}
& \langle \bar{B} P B \rangle : \\
&\frac{\Lambda}{\sqrt{6}} \left\{ \pi^0 \bar{\Sigma}^0 + \pi^+ \bar{\Sigma}^- + \pi^- \bar{\Sigma}^+ - \frac{\sqrt{2}}{3} \eta \bar{\Lambda} -2 K^+ \bar{p} -2 K^0 \bar{n} \right.\\
&\left.+ K^- \bar{\Xi}^+ + \bar{K}^0 \bar{\Xi}^0 \right\}.   
\end{aligned}
\end{equation}
The final state $ \Lambda \pi^+ \bar{\Sigma}^- $ appears in both structures of Eqs.~\eqref{eq:BBP} and~\eqref{eq:BPB}, but one can reach the same final state from final state interaction of all the other $ M \bar{B} $ states.

\begin{figure}[htbp]
	\centering
	
	%	\centering
	\includegraphics[scale=0.50]{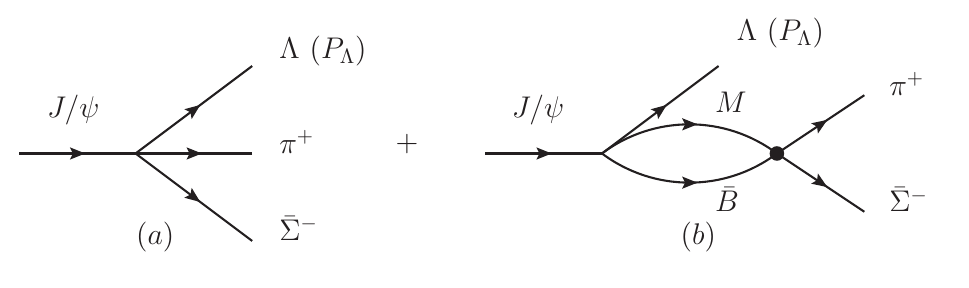}
	
	\caption{(a) Tree level in $ J/\psi \to \Lambda \pi^+ \bar{\Sigma}^- $ ; (b) rescattering of the $ M \bar{B} $ components.}\label{fig:new1}
\end{figure}

Before proceeding forward we recall now the structure of the $ J/\psi \to M B \bar{B} $ vertex of Eq.~\eqref{eq:t}, but now we should also take into account the structure with $ \epsilon_\mu p_\Lambda^\mu $. For symmetry reasons we take now the structure
\begin{equation}
 - [ \epsilon_\mu p_\Lambda^\mu + \epsilon_\mu p_{\bar{\Sigma}^-}^\mu ] ,
\end{equation}
which in the $ J/\psi$ rest frame will read as $ \vec{\epsilon} \cdot \vec{p}_\Lambda + \vec{\epsilon} \cdot \vec{p}_{\bar{\Sigma}^-} $. In the tree level both structures will contribute, but when we perform the final state interaction, as depicted in Fig.~\ref{fig:new1}, the $ p_{\bar{\Sigma}^-} $ ( $ p_{\bar{B}} $ in the loops ) terms will not contribute since they induce a $P$-wave vertex and the $ M \bar{B} \to \pi^+ \bar{\Sigma}^- $ amplitudes are in $S$-wave~\cite{Lyu:2026rsm}. As a consequence, from the structures of Eqs.~\eqref{eq:BBP} and~\eqref{eq:BPB} we shall get a matrix from $ J/\psi \to \Lambda \pi^+ \bar{\Sigma}^- $ given by (we remove the common factor $ 1/\sqrt{6} $),
\begin{equation}\label{eq:t_couple}
\begin{aligned} 
&\left(\frac{\tilde{A} + \tilde{B}}{2}\right) ( \vec{\epsilon} \cdot \vec{p}_\Lambda + \vec{\epsilon} \cdot \vec{p}_{\bar{\Sigma}^-} ) \\
&+ \vec{\epsilon} \cdot \vec{p}_\Lambda \left\{ \left(\frac{\tilde{A} + \tilde{B}}{2}\right) \left( G_{\pi^0 \Sigma^0} t_{\pi^0 \Sigma^0, \pi^- \Sigma^+} + G_{\pi^- \Sigma^+} t_{\pi^- \Sigma^+, \pi^- \Sigma^+} \right.\right.\\
&\left.+ G_{\pi^+ \Sigma^-} t_{\pi^+ \Sigma^-, \pi^- \Sigma^+} - \frac{\sqrt{2}}{3} G_{\eta \Lambda} t_{\eta \Lambda, \pi^- \Sigma^+} \right) \\
&+  \left(\frac{\tilde{A} - 2\tilde{B}}{2}\right)  \left( G_{K^+ p} t_{K^+ p, \pi^- \Sigma^+} + G_{K^0 n} t_{K^0 n, \pi^- \Sigma^+} \right) \\
& +  \left(\frac{-2\tilde{A} + \tilde{B}}{2}\right)  \left( G_{K^- \Xi^+} t_{K^- \Xi^+, \pi^- \Sigma^+}\right. \\
&\left.\left.+ G_{\bar{K}^0 \Xi^0} t_{\bar{K}^0 \Xi^0, \pi^- \Sigma^+} \right) \right\},
\end{aligned}
\end{equation}
where the argument of $ G $ and $ t $ is $ M_{\text{inv}}(\pi^+ \bar{\Sigma}^-) $, we have explicitly used that $ t_{M B, M B} = t_{\bar{M} \bar{B}, \bar{M} \bar{B}} $, in order to use the ordinary $ T $ matrices of the chiral unitary approach~\cite{Oset:1997it}, which are used in the present calculation. Note that the term $ \frac{\tilde{A} + \tilde{B}}{2} \vec{\epsilon} \cdot \vec{p}_{\bar{\Sigma}^-} $ in Eq.~\eqref{eq:t_couple} corresponds to the term with 1 in Eq.~\eqref{eq:a_and_b}.

Since there is an arbitrary normalization in the events of the data, we define
\begin{equation}
\frac{\tilde{A} + \tilde{B}}{2} = \tilde{C}, 
\end{equation}
and write the equations in terms of $ \tilde{C} $ and $ \tilde{A}/\tilde{B} $. If we sum the amplitudes of Eq.\eqref{eq:a_b_angle} coming from the $ \pi \Lambda $ interaction we can write the total amplitude as
\begin{equation}
t = t_1 \vec{\epsilon} \cdot \vec{p}_{\bar{\Sigma}^-} + t_2 \vec{\epsilon} \cdot \vec{\tilde{p}}_{\pi^+} + t_3 \vec{\epsilon} \cdot \vec{p}_\Lambda, 
\end{equation}
where
\begin{equation}\label{eq:t_3}
\begin{aligned}
&t_3 = \tilde{C} \left( 1 + G_{\pi^0 \Sigma^0} t_{\pi^0 \Sigma^0, \pi^- \Sigma+} + G_{\pi^- \Sigma^+} t_{\pi^- \Sigma^+, \pi^- \Sigma^+} \right.\\
&\left.+ G_{\pi^+ \Sigma^-} t_{\pi^+ \Sigma^-, \pi^- \Sigma^+} - \frac{\sqrt{2}}{3} G_{\eta \Lambda} t_{\eta \Lambda, \pi^- \Sigma^+} \right) \\
&+ \tilde{C}  \left(\frac{\tilde{A}/\tilde{B} - 2}{\tilde{A}/\tilde{B} + 1}\right)  \left( G_{K^+ p} t_{K^+ p, \pi^- \Sigma^+} + G_{K^0 n} t_{K^0 n, \pi^- \Sigma^+} \right) \\
& + \tilde{C}  \left(\frac{-2\tilde{A}/\tilde{B} + 1}{\tilde{A}/\tilde{B} + 1}\right) \left( G_{K^- \Xi^+} t_{K^- \Xi^+, \pi^- \Sigma^+} \right.\\
&\left.+ G_{\bar{K}^0 \Xi^0} t_{\bar{K}^0 \Xi^0, \pi^- \Sigma^+} \right). 
\end{aligned}
\end{equation}
\begin{equation}\label{eq:t_1}
\begin{aligned}
 t_1 =& \tilde{C} \left( 1 + \frac{\tilde{a}^\prime M_{\Sigma(1430)}}{M_{\pi \Lambda} - M_{\Sigma(1430)} + i \frac{\Gamma_{\Sigma(1430)}}{2}} \right.\\
 &\left.+ \frac{\tilde{a}_1^\prime M_{\Sigma(1380)}}{M_{\pi \Lambda} - M_{\Sigma(1380)} + i \frac{\Gamma_{\Sigma(1380)}}{2}} \right) .
\end{aligned}
\end{equation}
\begin{equation}\label{eq:t_2}
 t_2 = \frac{\tilde{b}^\prime M_{\Sigma(1385)}}{M_{\pi \Lambda} - M_{\Sigma(1385)} + i \frac{\Gamma_{\Sigma(1385)}}{2}}. 
\end{equation}
Note that for the $ \Sigma(1385) $ we have taken the term with $ \vec{\epsilon} \cdot \vec{\tilde{p}}_{\pi^+} $, where $ \vec{\tilde{p}}_{\pi^+} $ is the momentum of the pion in the $ \pi \Lambda $ rest frame, which is the $ \Sigma(1385) $ rest frame. Then we obtain
\begin{equation}\label{eq:t_sum}
\begin{aligned}
\bar{\sum} \sum |t|^2 =& \frac{1}{3} \left\{ |t_1|^2 \vec{p}_{\bar{\Sigma}^-}^{~2} + |t_2|^2 |\vec{\tilde{p}}_{\pi^+}|^{2} + |t_3|^2 \vec{p}_\Lambda^{~2} \right.\\
&+ 2 \text{Re}(t_1 t_2^*) \vec{p}_{\bar{\Sigma}^-} \cdot \vec{\tilde{p}}_{\pi^+} + 2 \text{Re}(t_1 t_3^*) \vec{p}_{\bar{\Sigma}^-} \cdot \vec{p}_\Lambda \\
&\left.+ 2 \text{Re}(t_2 t_3^*) \vec{\tilde{p}}_{\pi^+} \cdot \vec{p}_\Lambda \right\}. 
\end{aligned}
\end{equation}
We will now use Eq.~\eqref{eq:double_width} to calculate the mass distributions and we must write all scalar products in terms of invariant masses, which is easy, but to do it we must write $ \vec{\tilde{p}}_{\pi^+} $ in terms of momenta in the $J/\psi$ rest frame, where we have $ \vec{p}_\Lambda $ and $ \vec{p}_{\bar{\Sigma}^-} $. For this we use the boost from the frame of $ J/\psi$ at rest to the frame where $ \pi^+ \Lambda $ are at rest and taking into account that in the $ J/\psi$ rest frame $ \vec{p}_{\pi^+} + \vec{p}_\Lambda = -\vec{p}_{\bar{\Sigma}^-} $ and using Eq.~\eqref{eq:A4} of Ref.~\cite{Crdoba1995ProjectileDE}, we obtain:
\begin{equation}
\vec{\tilde{p}}_{\pi^+} = \left[ \left( \frac{E_{\pi \Lambda}}{M_{\pi \Lambda}} - 1 \right) \frac{\vec{p}_{\pi^+} \cdot \vec{p}_{\bar{\Sigma}^-}}{\vec{p}_{\bar{\Sigma}^-}^{~2}} + \frac{E_{\pi^+}}{M_{\pi \Lambda}} \right] \vec{p}_{\bar{\Sigma}^-} + \vec{p}_{\pi^+},
\end{equation}
where $ M_{\pi \Lambda} $ is the invariant mass of $ \pi \Lambda $,
\begin{equation}
E_{\pi \Lambda} = \sqrt{M_{\pi \Lambda}^2 + \vec{p}_{\bar{\Sigma}^-}^{~2}} \quad , \quad E_{\pi^+} = \frac{M_{J/\psi}^2 + m_{\pi^+}^2 - M_{\bar{\Sigma}^-\Lambda}^2}{2 M_{J/\psi}}. 
\end{equation}
And all the needed products of momenta are
\begin{equation}
 2 \vec{p}_{\pi^+} \cdot \vec{p}_{\bar{\Sigma}^-} = m_{\pi^+}^2 + m_{\bar{\Sigma}^-}^2 + 2 E_{\pi^+} E_{\bar{\Sigma}^-} - M_{\pi \bar{\Sigma}}^2, 
\end{equation}
with
\begin{equation}
  E_{\bar{\Sigma}^-} = \frac{m_{J/\psi}^2 + m_{\bar{\Sigma}^-}^2 - M_{\pi \Lambda}^2}{2 m_{J/\psi}},
\end{equation}
\begin{equation}
  2 \vec{p}_{\pi^+} \cdot \vec{p}_\Lambda = m_{\pi^+}^2 + m_\Lambda^2 + 2 E_{\pi^+} E_\Lambda - M_{\pi \Lambda}^2, 
\end{equation}
with
\begin{equation}
E_\Lambda = \frac{m_{J/\psi}^2 + m_\Lambda^2 - M_{\pi \bar{\Sigma}^-}^2}{2 m_{J/\psi}}, 
\end{equation}
\begin{equation}
 2 \vec{p}_\Lambda \cdot \vec{p}_{\bar{\Sigma}^-} = m_\Lambda^2 + m_{\bar{\Sigma}^-}^2 + 2 E_\Lambda E_{\bar{\Sigma}^-} - M_{\Lambda \bar{\Sigma}^-}^2.
\end{equation}
Everything is defined in terms of $ M_{12}, M_{13}, M_{23}$, and taking into account Eq.~\eqref{eq:M_relation}, all can be defined in terms of two invariant masses, $ M_{12}, M_{23} $. To calculate $ \Gamma $ or $ d\Gamma/dM_{12} $ ( $ M_{12} = M_{\pi \Lambda} $ ) one can use either of the formulas in Appendix~\ref{sec:appendix}, $ d^2\Gamma / dM_{12} d\cos\theta $ or $ d^2\Gamma / dM_{12} dM_{23} $. However, using $ d^2\Gamma / dM_{12} d\cos\theta $ we can see that there is no interference between $ t_1 $ and $ t_2 $ when integrating over $ \cos\theta $ to get $ d\Gamma/dM_{12} $ because $ t_1 $ depends on $ M_{12} $ and so does $ t_2 $, and
\begin{equation}
  \vec{\tilde{p}}_{\pi^+} \cdot \vec{p}_{\bar{\Sigma}^-} = \vec{\tilde{p}}_{\pi^+}  \cdot\vec{\tilde{p}}_{\bar{\Sigma}^-}\frac{M_{12}}{m_{J/\psi}} = |\vec{\tilde{p}}_{\pi^+}| |\vec{\tilde{p}}_{\bar{\Sigma}^-}|\frac{M_{12}}{m_{J/\psi}} \cos\theta ,
\end{equation}
with $ |\vec{\tilde{p}}_{\pi^+}|, |\vec{\tilde{p}}_{\bar{\Sigma}^-}| $ functions of $ M_{12} $ (see Appendix~\ref{sec:appendix}). Thus when integrating $ d^2\Gamma / dM_{12} d\cos\theta $ over $ \cos\theta $ the interference term $ 2 \text{Re}(t_1 t_2^*) \vec{\tilde{p}}_{\pi^+} \cdot \vec{p}_{\bar{\Sigma}^-} $ vanishes. However, one can not say the same about the interference term of $ t_2 $ and $ t_3 $ since $ t_3 $ depends on $ M_{23} $ and $ \vec{p}_\Lambda \cdot \vec{p}_{\bar{\Sigma}^-} $ depends on $ M_{13} $, $ \vec{p}_\Lambda $ on $ M_{23} $ and $ \vec{p}_{\bar{\Sigma}^-} $ on $ M_{12} $, and the term does not vanish upon integration over $ \cos\theta $. This interference is relevant to the conclusions of the paper.

The other lesson from this discussion is that now it is preferable to take the expression of $ d^2\Gamma / dM_{12} dM_{23} $ and integrate over $ M_{23} $ to obtain $ d\Gamma / dM_{12} $ since all terms in the final expression are easily written in terms of the three invariant masses.

\section{ Results with $\pi\bar{\Sigma}^-$ interaction}\label{sec5}

\begin{figure}[htbp]
	\centering
	
	%	\centering
	\includegraphics[scale=0.65]{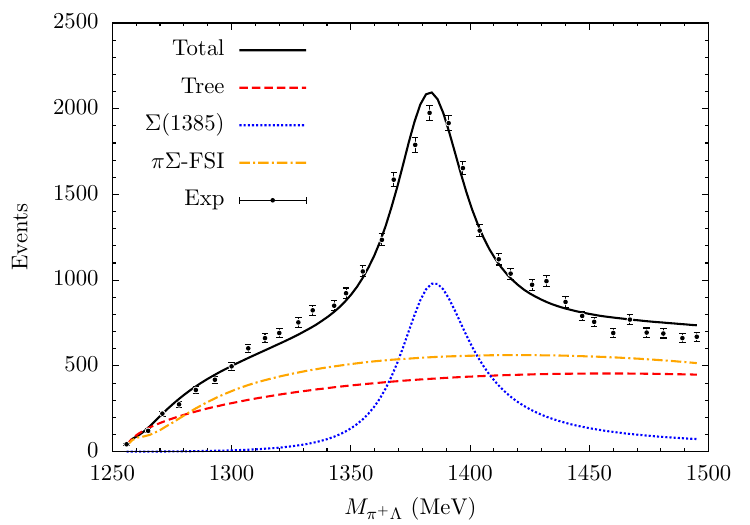}
	
	\caption{The $\pi^+\Lambda$ mass distribution considering the $\pi\Sigma$ interaction without contributions from $\Sigma(1430)$ and $\Sigma(1380)$. The data are taken from Ref.~\cite{BESIII:2023syz}.}\label{fig:new2}
\end{figure}
\begin{figure}[htbp]
	\centering
	
	%	\centering
	\includegraphics[scale=0.65]{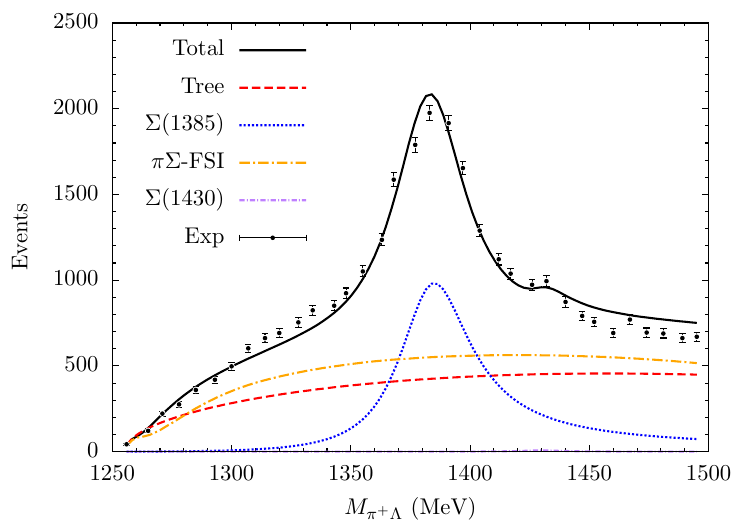}
	
	\caption{The $\pi^+\Lambda$ mass distribution considering the $\pi\Sigma$ interaction with contribution from $\Sigma(1430)$. The data are taken from Ref.~\cite{BESIII:2023syz}.}\label{fig:new3}
\end{figure}
In Fig.~\ref{fig:new2} we show the results that we obtain now, omitting the $ \Sigma(1380) $ and $ \Sigma(1430) $ contributions. We show in the figure the contribution of the $ |t_1|^2, |t_2|^2, |t_3|^2 $ terms together with the interference contributions. We have fitted the parameter $ \tilde{C} $ (global normalization) and the $ \tilde{b}^\prime $ parameter. We have chosen $ \tilde{A} = 3\tilde{B} $, in line with the results of Refs.~\cite{He:2026mkf,Ikeno:2026zpv}. The parameters used are $\tilde{C}=4.20\times10^{-3}$, $\tilde{b}^\prime=8.23\times10^{-2}$, $\tilde{a}^\prime=0$ and $\tilde{a}_1^\prime=0$. The results that we obtain are in good agreement with experiment and this was now made possible by taking into account the contribution of the $ \pi \bar{\Sigma}^- $ interaction including the interference terms. We do not aim at obtaining the other mass distributions, since this only distracts the attention from the main conclusion that we want to draw here, which is that in spite of the appealing results in the first part of the paper when considering only the $ \pi^+\Lambda $ interaction, which welcomed a contribution from the $ \Sigma(1380) $ resonance, the need for it has disappeared when considering at the same time the $ \pi\bar{\Sigma}^- $ interaction, together with other allowed coupled channels. While this says nothing about other works where the need for the contribution of the $ \Sigma(1380) $ is claimed, it serves as a warning that care must be taken when studying reactions claiming the need for the $ \Sigma(1380) $, since small contributions or subtle details can easily change the conclusions.

Coming back to the other invariant masses, we observe that the results depend very much on the value of $ \tilde{A}/\tilde{B} $, and in any case, contribution like the $ \bar{\Lambda}(1520)(3/2^+) $ in the $ \pi \bar{\Sigma}^- $ mass distribution are not included in our analysis. The $\bar{\Lambda}(1520)(3/2^+)$ appears in $D$-wave in the $\pi \bar{\Sigma}^-(1/2^-)$ mass distribution, and leads to a smooth background in the $\pi \Lambda$ mass distribution at low energies, easily accounted for by the tree level determined from the
fits of the $ \tilde{C}, \tilde{A}/\tilde{B} $ parameters. However, we observe that the results for the $ \pi^+\Lambda $ mass distribution are basically independent on the $ \tilde{A}/\tilde{B} $ ratio. Indeed, we have made calculations of the $M_{\pi\Lambda}$ distribution using values of $ \tilde{A}/\tilde{B} $ from 0.3 to 3 and in all cases we can get a very good fit to the data changing $ \tilde{C} $ and $ \tilde{b}^\prime $. Thus, our conclusion that the $ \Sigma(1380) $ state is not needed to explain the $ \pi \Lambda $ mass distribution is firm.

To finish with the subject we include now a contribution of the $ \Sigma(1430) $ and show the results in Fig.~\ref{fig:new3}. This is obtained using $\tilde{a}^{\prime}=9.0\times10^{-4}$. We see that the data welcome a small contribution, but it is based on a single data point. One obviously can not draw any conclusion from there, but only suggest that measurements with high statistics be performed to eventually show a clear signal of the $ \Sigma(1430) $, which we expect to be small, as already shown in the Belle experiment~\cite{Belle:2022ywa}.

\section{Quantitative discussion of the results}

\begin{table}
\centering
\caption{Values of $\chi^2$ and $\chi^2/d.o.f.$ for the different fit scenarios.}
\begin{tabular}{|c|c|c|c|c|c|}
	\hline
	~ & Fig. 2 & Fig. 3 & Fig. 4 & Fig. 6 & Fig. 7 \\
	\hline
	$\chi^2$ &~~$649.10$~~&~~$325.02$~~&~~$336.45$~~&~~$149.25$~~&~~$178.81$~~\\
    $\chi^2/d.o.f.$ &~~$19.67$~~&~~$10.16$~~&~~$10.85$~~&~~$4.52$~~&~~$5.59$~~\\
	\hline
\end{tabular}
\label{table:chi2_fit}
\end{table}

\begin{figure}[htbp]
	\centering	
	%	\centering
	\includegraphics[scale=0.65]{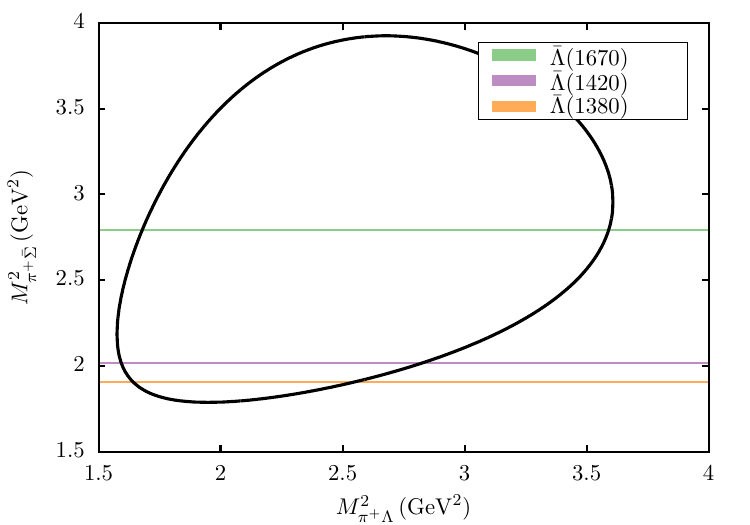}
	\caption{Dalitz plot in the 
    $M_{\pi^{+}\Lambda}^{2}$ --
    $M_{\pi^{+}\bar{\Sigma}}^{2}$ plane, with the colored bands indicating the relevant antihyperon resonances.}\label{fig:new4}
\end{figure}

We show here with numbers the improvements made from one fit to another. First, we compare the results of Figs.~\ref{fig:2},~\ref{fig:3},~\ref{fig:4}. In Table~\ref{table:chi2_fit} we show the values of $\chi^2$ and $\chi^2/\mathrm{d.o.f.}$ for the different scenarios. Note that in all cases the values of $\chi^2/\mathrm{d.o.f.}$ are relatively big, but this is due to the uncommon small errors of the data. Yet, inspection of the results in the figures, indicates that the fits in Figs.~\ref{fig:3} and~\ref{fig:4} are very good for a theoretical model, in spite of having values of $\chi^2/\mathrm{d.o.f.}$ of the order of 10. Hence, what is significative for us is the relative values of the $\chi^2/\mathrm{d.o.f.}$ for the different scenarios. Then we can see that by including the $\Sigma(1380)$ contribution $\chi^2/\mathrm{d.o.f.}$ is reduced by a factor of two, going from Fig.~\ref{fig:2} to Fig.~\ref{fig:3}. Going from Fig.~\ref{fig:3} to Fig.~\ref{fig:4}, introducing a small contribution from the $\Sigma(1430)$, there is a better agreement with the data in the region of $1430$~MeV, but the small change has not produced an improvement of the global $\chi^2/\mathrm{d.o.f.}$ value.

On the other hand, by removing the $\Sigma(1380)$ contribution and considering the $\pi\bar{\Sigma}$ interaction, we obtain Fig.~\ref{fig:new2} for which the $\chi^2/\mathrm{d.o.f.}$ has been reduced by more than a factor of two with respect to Fig.~\ref{fig:3}, indicating a considerable improvement in the fit to the data. We also see that trying to fit the $1430$~MeV region introducing a small $\Sigma(1430)$ contribution in Fig.~\ref{fig:new3}, improves this energy region at the expense of a somewhat worse agreement at higher invariant masses, that makes the $\chi^2/\mathrm{d.o.f.}$ value increase a little with respect to the fit of Fig.~\ref{fig:new2}, but it is still about a factor of two smaller than the corresponding fit in Fig.~\ref{fig:4}.

The conclusion of our study is that the fit to the data considering the $\pi\bar{\Sigma}$ interaction is better than that obtained by neglecting this interaction and introducing the $\Sigma(1380)$ contribution. This, together with the fact that the $\pi\bar{\Sigma}$ interaction can not be neglected, indicates that the present reaction cannot be used to check the need for an eventual $\Sigma(1380)$ contribution.

Finally, we would like to show why the $\pi\bar{\Sigma}$ interaction has affected substantially the $\pi\Lambda$ mass distribution. For that, we show in Fig.~\ref{fig:new4} the Dalitz plot and the resonances that the $\pi\bar{\Sigma}$, together with its coupled channels, generate. There are the $\bar{\Lambda}(1380)$, $\bar{\Lambda}(1420)$, $\bar{\Lambda}(1670)$. The region of $M_{\pi\Lambda}$ analysed in figures goes up to $M_{\pi\Lambda}^{2}=2.25\ \mathrm{GeV}^{2}$ and we can see that all these resonances in the $\pi\bar{\Sigma}$ influence the $\pi\Lambda$ mass distribution.

We also want to add another comment. As one can see in the detailed contributions of different terms in the figures, the contribution of the $\Sigma(1380)$ term by itself is extremely small, as can be seen in Figs.~\ref{fig:3} and~\ref{fig:4}. So is the contribution of the $\Sigma(1430)$ in Figs.~\ref{fig:4} and~\ref{fig:new3}. It is the interference with other terms that make their contributions relatively sizeable. However, the contribution of the $\pi\bar{\Sigma}$ final interaction by itself is quite large, as can be seen in Figs.~\ref{fig:new2} and~\ref{fig:new3}. The good fit obtained introducing this contribution is due to a smaller size of the $\Sigma(1385)$ contribution.

\section{ Conclusions }\label{sec6}
We studied the $J/\psi \to \Lambda \pi \bar{\Sigma}$ reaction measured by the BESIII Collaboration, looking at the $\pi^+ \Lambda$ mass distribution at low energies, where  $\Sigma$ states should show up. Indeed, there is a big peak in the mass distribution which corresponds to the $\Sigma(1385)$ excitation, but we took advantage to see if there would also be some signal for the $\Sigma(1430)$, predicted by works using the chiral unitary approach in coupled channels, and already observed experimentally in a Belle experiment. We also looked for a possible signal of the resonance $\Sigma(1380)$, claimed by many works from analysis of different reactions, and predicted as a pentaquark state in some quark model works.

A first analysis of the data using only structures in the invariant mass of $\pi^+ \Lambda$ shows that the low energy part of the $\pi^+ \Lambda$ spectrum demands the contribution of the $\Sigma(1380)$, but also there is room for a smaller contribution of the $\Sigma(1430)$. 
  
In a second step we look at the double differential mass distribution by considering the interaction of $\pi \bar{\Sigma}$ in order to see if this consideration can fill the missing strength in the $\pi^+ \Lambda$ mass distribution at low energies.  

What we see after taking into account the $\pi \bar{\Sigma}$ interaction, together with that of other possible coupled channels, there is no longer need for the $\Sigma(1380)$ contribution. The terms coming from the $\pi \bar{\Sigma}$ interaction, together with that of other possible coupled channels, provide a non-negligible contribution to the $\pi \Lambda$ mass distribution and also interfere with the dominant $\Sigma(1385)$ contribution, such that now a good reproduction of the data is obtained. It is also rewarding to see that the results obtained are independent on the ratio of two parameters of the theory and, up to a global normalization, only depend on the strength of the $\Sigma(1385)$ term, which makes the conclusions rather solid.

In summary we show that the present reaction cannot be used to claim the need for a $\Sigma(1380)$ resonance. Nothing can be said about other reactions from where it has been claimed. However, the exercise done here shows that a careful analysis of data is needed, and subtle issues or small contributions can change the conclusions concerning the need of that resonance to explain certain experimental data.

\section*{Acknowledgments}
This work was supported by the National Key R\&D Program of China (Grant No. 2024YFE0105200), the National Natural Science Foundation of China under Grant No. 12475086 and No. 12192263, the Natural Science Foundation of Henan (Grant No. 252300423951), and the Zhengzhou University Young Student Basic Research Projects for PhD students (Grant No. ZDBJ202522). Wen-Tao Lyu acknowledges the support of the China Scholarship Council. 
This work is also partly supported by the Spanish Ministerio de Economia y Competitividad~(MINECO) and European FEDER funds under Contracts No. FIS2017-84038-C2-1-PB, PID2020-112777GB-I00, and by Generalitat Valenciana under contract PROMETEO/2020/023. This project has received funding from the European Union Horizon 2020 research and innovation program under the program H2020-INFRAIA-2018-1, grant agreement No. 824093 of the STRONG-2020 project.

\appendix
\section{Relationship of $ \frac{d^2\Gamma}{dM_{12} dM_{23}} $ and $ \frac{d^2\Gamma}{dM_{12} d\cos\theta} $}\label{sec:appendix}
\begin{figure}[htbp]
	\centering
	
	%	\centering
	\includegraphics[scale=0.65]{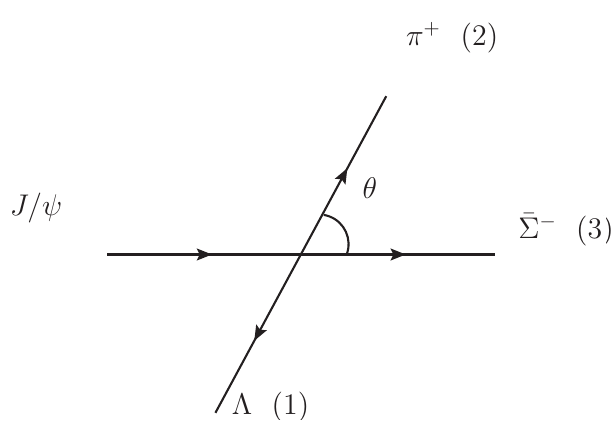}
	
	\caption{Particles in $ J/\psi \to \Lambda \pi^+ \bar{\Sigma}^- $ in the 1, 2 rest frame.}\label{fig:A1}
\end{figure}

Let us consider the three particles $ \Lambda(1), \pi^+(2), \bar{\Sigma}^-(3) $ in $ J/\psi\to1+2+3 $. In the case of three particles and sum of polarizations $ d\Gamma $ depends on two variables $ M_{12}, M_{23} $~\cite{ParticleDataGroup:2024cfk}, and $ M_{13} $ is given in terms of them:
\begin{equation}
  M_{12}^2 + M_{13}^2 + M_{23}^2 = M_{J/\psi}^2 + m_1^2 + m_2^2 + m_3^2. 
\end{equation}
One can write~\cite{ParticleDataGroup:2024cfk} (including the $ 2m_{\Lambda} 2M_{\bar{\Sigma}^-} $ normalization factors in $ \overline{\sum} \sum |t|^2 $ for simplicity)
\begin{equation}
\frac{d^2\Gamma}{dM_{12}^2 dM_{23}^2} = \frac{1}{(2\pi)^3} \frac{1}{32 m_{J/\psi}^3} \overline{\sum} \sum |t|^2.
\end{equation}
Let us work in the 1, 2 rest frame ($ \pi^+ \Lambda $) depicted in Fig.~\ref{fig:A1}.
We call $ \vec{p}_i $ the momenta in the $J/\psi$ rest frame and $ \vec{\tilde{p}}_i $ the momenta in the 1, 2 rest frame. Note that we go from the $J/\psi$ rest frame to the 1, 2 rest frame by making a boost along the direction of $ \vec{p}_{\bar{\Sigma}^-}(3) $ and opposite sign. Note also that in the 1, 2 rest frame
\begin{equation}
\vec{\tilde{p}}_{J/\psi} = \vec{\tilde{p}}_{\bar{\Sigma}^-}.
\end{equation}
We define as $ \theta $ angle the one of particle 2 ($ \pi^+ $) with particle 3 ($ \bar{\Sigma}^- $) in the 1, 2 rest frame. Note that this angle is the same as the one of particle 2 ($ \pi^+ $) in the 1, 2 rest frame with the momentum $ \vec{p}_{\bar{\Sigma}^-} $ in the $J/\psi$ rest frame, since $ \vec{\tilde{p}}_{\bar{\Sigma}^-} $ and $ \vec{p}_{\bar{\Sigma}^-} $ are parallel.

Let us calculate $ M_{23}^2 $ in the 1, 2 rest frame.
\begin{equation}\label{eq:A4}
M_{23}^2 = (p_2 + p_3)^2 = m_2^2 + m_3^2 + 2 \tilde{E}_2 \tilde{E}_3 - 2 \vec{\tilde{p}}_2 \cdot \vec{\tilde{p}}_3,
\end{equation}
where
\begin{equation}\label{eq:A5}
\tilde{E}_2 = \frac{M_{12}^2 + m_2^2 - m_1^2}{2 M_{12}},
\end{equation}
\begin{equation}
\sqrt{M_{J/\psi}^2 +\vec{\tilde{p}}_3^{~2}} = M_{12}+\sqrt{m_3^2+\vec{\tilde{p}}_3^{~2}},
\end{equation}
\begin{equation}\label{eq:A6}
M_{J/\psi}^2 +\vec{\tilde{p}}_3^{~2}= M_{12}^2 + m_3^2 +\vec{\tilde{p}}_3^{~2} + 2 M_{12} \tilde{E}_3,
\end{equation}
from where
\begin{equation}\label{eq:A7}
\tilde{E}_3 = \frac{M_{J/\psi}^2 - M_{12}^2 - m_3^2}{2 M_{12}},
\end{equation}
and
\begin{equation}\label{eq:A8}
\tilde{p}_2 = \frac{\lambda^{1/2}(M_{12}^2, m_1^2, m_2^2)}{2 M_{12}},
\end{equation}
\begin{equation}\label{eq:A9}
\tilde{p}_3 = \frac{\lambda^{1/2}(M_{J/\psi}^2, m_3^2, M_{12}^2)}{2 M_{12}}.
\end{equation}
Note that
\begin{equation}\label{eq:A10}
p_3 = \frac{\lambda^{1/2}(m_{J/\psi}^2, m_3^2, M_{12}^2)}{2 m_{J/\psi}}.
\end{equation}
Then Eq.~\eqref{eq:A4} gives
\begin{equation}\label{eq:A11}
M_{23}^2 = m_2^2 + m_3^2 + 2 \tilde{E}_2 \tilde{E}_3 - 2 |\vec{\tilde{p}}_2| |\vec{\tilde{p}}_3| \cos\theta,
\end{equation}
and $ \tilde{E}_2, \tilde{E}_3, \tilde{p}_2, \tilde{p}_3 $ all depend exclusively on $ M_{12} $, and not on the other two invariant masses. Then we have
\begin{equation}\label{eq:A12}
\Gamma = \int \frac{d^2\Gamma}{dM_{12}^2 dM_{23}^2} dM_{12}^2 dM_{23}^2,
\end{equation}
and we can make a change of variables to $ M_{12}, \cos\theta $. The Jacobian is given by
\begin{equation}
\label{eq:A13}
\begin{vmatrix}
\dfrac{\partial M_{12}^2}{\partial M_{12}^2} &
\dfrac{\partial M_{12}^2}{\partial \cos\theta}
\\[1.5ex]
\dfrac{\partial M_{23}^2}{\partial M_{12}^2} &
\dfrac{\partial M_{23}^2}{\partial \cos\theta}
\end{vmatrix}
=
\begin{vmatrix}
1 & 0
\\[1.5ex]
0 & -2\tilde{p}_2\tilde{p}_3
\end{vmatrix}.
\end{equation}
Hence, we can write
\begin{equation}\label{eq:A14}
\begin{aligned}
\Gamma &= \int \frac{d^2\Gamma}{dM_{12}^2 dM_{23}^2} 2 \tilde{p}_2 \tilde{p}_3 dM_{12}^2 d\cos\theta \\
&= \int \frac{1}{(2\pi)^3} \frac{1}{32 m_{J/\psi}^3} \bar{\sum} \sum |t|^2 2 \tilde{p}_2 \tilde{p}_3 2 M_{12} dM_{12} d\cos\theta \quad.
\end{aligned}
\end{equation}
And using Eqs.~\eqref{eq:A9} and~\eqref{eq:A10} we can write
\begin{equation}\label{eq:A15}
\frac{d\Gamma}{dM_{12}} = \int \frac{1}{(2\pi)^3} \frac{1}{8 m_{J/\psi}^2} p_3 \tilde{p}_2 \overline{\sum} \sum |t|^2 d\cos\theta,
\end{equation}
and equivalently
\begin{equation}\label{eq:A16}
\frac{d^2\Gamma}{dM_{12} d\tilde{\Omega}} = \frac{1}{(2\pi)^4} \frac{1}{8 m_{J/\psi}^2} p_3 \tilde{p}_2 \overline{\sum} \sum |t|^2,
\end{equation}
with $ \tilde{\Omega} $ the solid angle in the 1, 2 rest frame. Eqs.~\eqref{eq:A11} and~\eqref{eq:A14} can be used to evaluate the decay width of a particle. In the case that the $ t $ matrix depends only on the variables of the particles 1, 2, Eq.~\eqref{eq:A15} is convenient because one can make a separation in partial waves, and different partial waves do not interfere in $ \frac{d\Gamma}{dM_{\text{inv}}} $. However, if the $ t $ matrix depends for instance on $ M_{12} $ and $ M_{23} $, there can be interference between different partial waves in the 1, 2 variables and those in the 2, 3 variables, because $ M_{23} $ depends on $ \cos\theta $ (see Eq.~\eqref{eq:A4}). In this case, using Eq.~\eqref{eq:A12} with the sum of the total $ t $ matrix is more convenient.

\end{document}